\documentclass[letter,twocolumn]{jpsj5}
\usepackage{bm}
\usepackage{graphicx,color}
\usepackage{color}
\begin{document}

\title{
Composite Orders and Lifshitz Transition of Heavy Electrons
}

\author{
Yoshio Kuramoto and Shintaro Hoshino$^1$
}
\inst{
Department of Physics, Tohoku University, Sendai 980-8578, Japan\\
$^1$Department of Basic Science, The University of Tokyo, Meguro, Tokyo 153-8902, Japan
}

\recdate{\today}

\abst{
Magnetic and other unconventional electronic orders are discussed for heavy electrons. 
In addition to the ordinary Kondo lattice, we consider non-Kramers systems taking the two-channel Kondo lattice, and another lattice that consists of a singlet-triplet $f^2$ configuration.  The latter gives a scalar order with
staggered sublattices of crystalline-electric-field (CEF) and Kondo singlets, which accounts for the observed order in PrFe$_4$P$_{12}$.
In the two-channel Kondo lattice, an odd-frequency pairing is realized that
becomes degenerate at half-filling
with a composite order involving both localized and conduction electrons.
The degeneracy is interpreted in terms of a hidden SO(5) symmetry. 
For the ordinary Kondo lattice with exchange interaction $J$, we take
insulating ground states as reference.
The metallic states are approached by 
infinitesimal doping of carriers
either from the Kondo insulator or from the antiferromagnetic insulator. 
By adding intersite exchange 
we find that location of the band extremum of the heavy band 
depends on the value of $J$.
The change of $J$ gives topological change of the Fermi surface, 
which corresponds to a generalized version of the Lifshitz transition, and
occurs separately from the quantum critical point of the antiferromagnetic order.
}

\kword{Kondo insulator, Lifshitz transition, quantum critical point,
two-channel Kondo lattice, composite order, odd-frequency order, SO(5) symmetry}

\maketitle

\section{Introduction}

This paper discusses peculiar
orderings of heavy electrons
on the basis of our recent theoretical results.
We distinguish the heavy-electron systems according to the number of $f$ electrons per ion.  
In cases of trivalent Ce and Yb systems,  there are single $f$-electron (Ce) or single hole (Yb) per ion.  These are called
Kramers systems.
On the other hand, trivalent Pr and tetravalent U systems have two $f$ electrons per site.  These are called
non-Kramers systems to which the Kramers theorem on 
the time reversal degeneracy does not apply.
The nature of an electronic order depends heavily on whether we are dealing with Kramers or non-Kramers systems, as we shall discuss in detail below.

The key theoretical framework is the dynamical mean-field theory (DMFT) 
that reduces the periodic lattice problem to that 
of the effective single-impurity Kondo system\cite{kuramoto85,georges96}.
Because of 
innovations in numerical tools such as the continuous-time quantum Monte Carlo \cite{rubtsov04,otsuki07,gull11} and the numerical renormalization group\cite{wilson75,bulla}, we can
now approach the ordering phenomena of heavy electrons.
At the quantum critical point (QCP),  the 
N\'{e}el temperature $T_N$ of the antiferromagnetic (AFM) order vanishes, by application of pressure for example.
Active discussion is going on whether the AFM QCP is related to the itinerant-localized transition of heavy electrons \cite{gegenwart}.  
Closely related issue is whether Kondo effect
breaks down by formation of the AFM order.
In particular, the parent (``normal") state which 
deviates from the Fermi-liquid behavior is attracting great attention\cite{lohneysen2007}.  
We take an option to start from the Kondo insulator 
that also undergoes a transition to an AFM order.  
Approaching the metallic heavy-electron system by doping the Kondo insulator, we can make qualitative but definite statement on the nature of the Fermi surface. 

Depending on the strength of hybridization with conduction electrons, either itinerant or localized picture becomes the appropriate starting point.  
Even though we start from the localized limit, it is possible to move to the itinerant picture by including the Kondo effect.
This crossover has been demonstrated for the Kondo lattice with a single localized electron per site \cite{otsuki09}.

Such crossover can take place also in non-Kramers systems with more variety.  First, let us assume the presence of multiple conduction bands, and
regard a non-Kramers system as a double Kondo lattice with a local $f^2$ configuration  
composed of spin singlet and triplet. 
If the $f^2$ singlet is lower than the triplet, we can identify the singlet as a crystalline-electric-field (CEF) singlet, and the triplet as the excited CEF level.
Then the non-Kramers $f$-electron state is naturally described.  
We have demonstrated \cite{hoshino2010,kuramoto2011}, in the borderline case as in PrFe$_4$P$_{12}$, an exotic scalar order can be realized that consists of alternating sites of Kondo and CEF singlets.

Secondly, we take the case where the CEF ground state is a non-Kramers doublet. 
In this case
the two-channel Kondo effect arise in the single-impurity case with the non Fermi-liquid ground state \cite{cox87, cox98}.  
The two-channel Kondo lattice cannot remain disordered  
in the ground state with intersite interaction.
We argue that 
strange diagonal and off-diagonal orders can emerge that involve both itinerant and localized parts of electrons\cite{hoshino11, hoshino13, hoshino2013}.
Especially such composite order includes a superconductivity which emerges directly from the non Fermi-liquid state.
In this paper we pay particular attention to the symmetry aspect of the composite order, which in fact is equivalent to an odd-frequency order. 

\section{Toy Model Composed of Four Spins}

We regard the Kondo insulator and a variant of the Mott insulator from a common perspective.  It is convenient to consider a case where each site of the Mott insulator has two spins, instead of one.   
Then we can take a nonmagnetic Mott insulator since the two spins can form a singlet at each site. The electrons in the Mott insulator are regarded as localized.  
On the other hand, the conduction electrons in the Kondo insulator are like those in the band insulator.  

\begin{figure}
\begin{center}
\includegraphics[width= 0.25\textwidth]{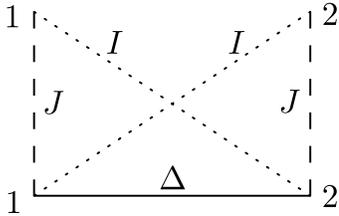}
\caption{
Illustration of exchange interactions in the four-spin system where each vertex labeled either 1 or 2 has a spin. The interaction 
$\Delta$ works between two $f$-electron spins, while $J$ and $I$ work between
$f$- and $c$-electron spins.
}\label{4spins}
\end{center}
\end{figure}

Let us understand how the distinction appears taking the simplest possible case in the strong-coupling limit.
If the adiabatic continuity is present, we can extrapolate the distinction to weaker coupling, or a large width of conduction bands.
We take two spins 
$\bm S_1$ and $\bm S_2$ 
representing $f$ electrons, and other two spins 
$\bm s_1$ and $\bm s_2$ 
representing conduction electrons.  The kinetic energy of the latter is neglected for the moment.
Namely we consider the model:
\begin{align}
{\cal H}_4 &= 
\Delta \bm S_1\cdot \bm S_2 + 
J(\bm S_1\cdot \bm s_1 +\bm S_2\cdot \bm s_2 )\nonumber\\
&+ I(\bm S_1\cdot \bm s_2 + \bm S_2\cdot \bm s_1 ),
\label{4spin-model}
\end{align}
as illustrated in Fig.\ref{4spins}. 
If ${\cal H}_4$ represents a single-site model, the parameter $\Delta\ (>0)$ can simulate the singlet-triplet splitting in the non-Kramers CEF states.   
If, on the other hand, ${\cal H}_4$ is meant for two sites,
$\Delta$ represents the AF Heisenberg coupling between neighboring $f$ spins.
In either case the parameter $J \ (>0)$ represents AF exchange between local and conduction electrons at the same site, while 
$I\ (>0)$ is another AF exchange, which acts 
with the neighboring site in the Kramers case.
In the non-Kramers case $I$ acts
with different conduction orbitals at the same site.

%%%%%%%%%%%%%
%%%%%%%%%%%%%

This model can be solved analytically for arbitrary value of  $I, J, \Delta$ by exploiting the parity of states.\cite{rasul89} 
We take here an alternative approach focusing on hidden symmetries.
Taking $I=0$ for the moment, we consider the two opposite limits. \\
(i) $J\gg \Delta$: 
The ground state in this case has two singlet pairs (2KS) of $f$- and $c$-electrons, which is a prototype of the Kondo insulator.  
The energy is $E_{2KS} =-3J/2$. \\
(ii) $\Delta \gg J$:
The $f$ spins forms a pair singlet (PS) in the ground state, and $c$ electrons are non-interacting. 
The PS is a prototype of the Mott insulator.  The energy becomes of $E_{PS} =-3\Delta/4$. \\
As $\Delta/J$ increases, (i) and (ii) are smoothly connected. 
Namely, the ground state in $\Delta \sim 2J$ consists basically of 
a superposition of  2KS and PS states. 

For $I\neq 0$, we consider
a special case $I=J$, 
where a new conserved quantity 
$\bm S\equiv \bm S_1+\bm S_2$ (or equivalently $\bm s \equiv\bm s_1+\bm s_2$) appears in addition to the total spin $\bm T\equiv \bm S+\bm s$.  
Hence the symmetry of the system is elevated to SU(2)$\otimes$SU(2) that is isomorphic to SO(4).
As a result, new degeneracies arise among different values of the total spin $T$.
The emergent higher symmetry simulates more complicated ones in large systems. Since the simplest case cultivates intuition,  we analyze the symmetry in more detail below.

Let us rewrite the Hamiltonian with $I=J$ as
\begin{align}
{\cal H}_4 = \frac 
\Delta 2
\left(   \bm S^2 - \frac 32
\right) +\frac 
J2 \left(\bm T^2 -\bm S^2-\bm s^2 \right).
\label{I=J}
\end{align}
Then it becomes manifest that eigenvalues can be
labeled by the quantum numbers $T\ (\le 2), S\ (\le 1)$ and $s\ (\le 1)$ as $E_{TSs}$. 
All 16 states are precisely given by possible combination of the quantum numbers with the triangular condition $|S-s|\le T\le |S+s|$.
It is obvious from eq.(\ref{I=J}) that we have
$E_{000} =E_{101}= -3\Delta/4$,
which means that the PS
is degenerate with a triplet. 
This four-fold degeneracy is a consequence of the SO(4) symmetry, which is analogous to that in hydrogen atom where all states for a given principal quantum number
are degenerate even with different values of angular momentum. 

Let us ask how the ground state changes from PS as $J/\Delta$ increases from zero under the condition $I=J$.
Instead of the smooth crossover to 2KS, we now have the critical state
in the special case of $\Delta=2J$.  In this case we obtain:
\begin{align}
{\cal H}_4 = J(
\bm T\cdot \bm S-3/4),
\end{align}
where the energy $-3J/4$ with $S=0$ does not depend on the value of $T$, and 
the energy with $T=0$ does not depend on $S$.
We then obtain  
$E_{011}=E_{000}=E_{101}$.
Namely the ground state is five-fold degenerate with two singlets and one triplet.  
The degeneracy of two singlets caused by a hidden symmetry  persists to the case of finite width of the conduction bands, as far as the particle-hole symmetry is preserved\cite{affleck92-2,bulla}.  
As a consequence, a local non-Fermi liquid is realized in the ground state.  It is known that the fixed point of the two-site model shares the same property as that in the two-channel Kondo impurity \cite{affleck92-2}.

\section{Unconventional Orders in Non-Kramers Systems}
\subsection{Orbital Kondo Effect}
The $f$ electron number per ion is two in Pr$^{3+}$ and U$^{4+}$,
and the Kramers theorem about the time-reversal degeneracy does not apply.
These ions are therefore called non-Kramers ions. 
The CEF ground state of a non-Kramers ion can either be a singlet, or a multiplet.
Those systems such as
PrMg$_3$ \cite{tanida06}, PrAg$_2$In \cite{yatskar96}, PrPb$_3$\cite{onimaru} and PrV$_2$Al$_{20}$\cite{sakai} 
have doubly degenerate CEF states as the ground state.
In this section we discuss electronic orders  
caused by interaction of the non-Kramers $f$ electrons with conduction electrons.

Since the spatial distribution of a wave function in the doubly degenerate state
is different from each other in general, quadrupole degree of freedom should arise.
In addition, an imaginary coefficient in 
linear combination of wave functions brings about
magnetic degree of freedom such as octupoles ($2^3$) and triakontadipoles ($2^5$).

We use
pseudospin $\hat S_\alpha\ (\alpha=x,y,z)$ 
to describe transitions inside the non-Kramers doublet.
The corresponding orbital degree of freedom of 
conduction electrons is also simulated by pseudospins.
Since a conduction electron has the real spin degrees of freedom as well, the number of 
combined degrees of freedom (orbit and spin) are more than 
that of the localized degree of freedom (orbit only) to be screened.
This situation has long been studied as the two-channel Kondo problem. \cite{cox98} 
It is known that the ground state including a non-Kramers impurity has a finite entropy\cite{cox98}, and is not a local Fermi liquid in contrast with spin Kondo systems\cite{nozieres80}.

On the other hand, if the CEF ground state is a singlet with small splitting below a triplet, 
the spin Kondo effect occurs even for non-Kramers systems.  The situation is just as illustrated in Fig.\ref{4spins} where each electron of $f^2$ configuration has the AFM interaction $J$ with conduction electrons, and the CEF splitting is given by $\Delta$.

Both spin and orbital types of Kondo effect bring about 
logarithmic temperature dependence $\ln T$ of the electrical resistivity $\rho(T)$ at high temperatures.
However,  clear examples of $\ln T$ behavior caused by the orbital Kondo effect are not abundant. 
The best candidate is
{\color[rgb]{0.000000,0.000000,0.000000}UBe}$_{13}$ where $\rho(T)$ increases with decreasing $T$, and saturates around $T\sim 40$ K\cite{ott83}.
Superconductivity emerges below about 1 K from a non-Fermi liquid state.
The CEF doublet $\Gamma_3$ system
PrV$_2$Al$_{10}$ 
also shows $\ln T$ behavior \cite{sakai}  
at $T\simeq 200$K.

On the other hand, Kondo effect is not seen in resistivity in such systems as PrMg$_3$ and PrAg$_2$In, which 
do not even show the symptom of electronic order at least down to 0.1  K. 
As a result, the $T$-coefficient of the specific heat becomes huge, reaching to several J/(mole$\cdot$K$^2)$.
One possible reason for the absence of Kondo effect is ferro-coupling of $f$ and $c$ quadruples $J_{\rm Q}$.
It is desirable to evaluate $J_{\rm Q}$ for non-Kramers doublet systems taking realistic  electronic structure. 

\subsection{Scalar Staggered Order}

We consider the following model for a non-Kramers system with  singlet and triplet CEF levels:
\begin{align}
{\cal H} & = 
   \sum _{\mib{k}\sigma} 
   \sum _{\alpha=1,2} 
 (\varepsilon _{\mib{k}\alpha} -\mu)
 c_{\mib{k}\alpha\sigma} ^\dagger c_{\mib{k}\alpha\sigma}
 + 
 J \sum _{i \alpha} \mib{S}_{\alpha i} \cdot \mib{s}_{\alpha i}    \nonumber \\
 & + 
\Delta  \sum_{i}\mib{S}_{1i} \cdot \mib{S}_{2i},
\label{singlet-triplet-model}
\end{align}
where 
$\mib{s}_{\alpha i}$ denotes the conduction electron spin at site $i$ with orbital $\alpha$.
The CEF splitting $\Delta$ at each site is simulated by
the coupling between local spins
$\mib{S}_{1i}$ and $\mib{S}_{2i}$
 as shown in the last term.

If we neglect the band-width, and put two conduction electrons at each site, the situation is expressed by eq.(\ref{4spin-model}) with $I=0$.
With finite band-width, 
the 2KS state of Fig.\ref{4spins} is connected with two-band Kondo insulator. 
On the other hand, the PS state corresponds to the singlet CEF state.
The homogeneous array of PS should be a metal since two conduction bands are both half-filled.

W now consider the quarter-filled case with one electron per site on the average.
Appropriate choice of chemical potential $\mu$ in eq.(\ref{singlet-triplet-model}) realizes the situation.
When the parameter $J/\Delta$ is increased from 0,  there must be a change from the singlet CEF lattice to the state involving Kondo singlets. 
It has been found \cite{hoshino2010} that a staggered state with neighboring CEF and Kondo singlets is realized for $\Delta/J\lesssim 0.4$ below the transition temperature $T_{\rm c}$ which is of the order of the Kondo temperature $T_{\rm K}$ and is much smaller than $J, \Delta$.  
Note that this electronic order does not break the point-group symmetry at each site, and hence is called a scalar order.
The scalar order accompanies the CDW of conduction electrons since the Kondo singlet site gathers more conduction electrons for the screening.
With quarter-filling
of two identical conduction bands, the ground state with the staggered scalar is an insulator.
We can also regard the state as a mixture of Kondo and nonmagnetic Mott insulators where each insulating state selects half of the whole lattice sites.

With generalization to non-identical conduction bands, the model seems to apply to the mysterious scalar order found in PrFe$_4$P$_{12}$\cite{hoshino2010}.  
Namely, in spite of the sharp peak of the magnetic susceptibility at the transition temperature
$T_c\sim 7$ K, no magnetic moment was found in the system \cite{torikachvili87}.  Then quadrupole order was invoked for some time \cite{hao03}.   Detailed diffraction and NMR measurements, however, found no deviation from the cubic symmetry.
The model calculation assuming the scalar order 
does indeed reproduced the peak in the magnetic susceptibility \cite{hoshino2010}.  The remarkable difference from the AFM is that the susceptibility remains isotropic in the ordered phase \cite{matsuda02}.

\begin{figure}
\begin{center}
\includegraphics[width= 0.4\textwidth]{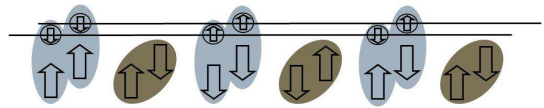}
\end{center}
\caption{(Color online)
Staggered order of CEF and Kondo singlets with a scalar order parameter.
}
\label{PrFeP}\end{figure}

Figure \ref{PrFeP} illustrates the scalar order in the strong-coupling limit.
In the actual case, the Kondo clouds extend well to CEF sites.  The staggered order should be probed by modulation of CEF splittings, or by spin correlations such as $\langle \bm S_i\cdot \bm s_i\rangle$ averaged over orbitals.  
Although the CDW of conduction electrons is also an order parameter, the essence of the order is on the $f$-electron part.
The scalar order is rather fragile in the model calculation, and requires
a restrictive condition for the magnitudes of parameters.
Experimentally,  the scalar order is indeed destroyed by alloying \cite{tayama07} or applying pressure \cite{hidaka05}.

%%%%%%%%%%%%%
%%%%%%%%%%%%%

\subsection{Diagonal Order of Itinerant-Localized Composites}

We now turn to the two-channel Kondo lattice (2chKL) as a model of systems of non-Kramers doublets coupled with conduction electrons.
The Hamiltonian is given by
\cite{jarrell96}
\begin{align}
{\cal H} = 
\sum _{\bm{k}\sigma} \sum_{\alpha=1,2}
 \varepsilon _{\bm{k}} c_{\bm{k}\alpha \sigma}^\dagger c_{\bm{k}\alpha \sigma}
+ J \sum_{i\alpha} \hat{\bm S}_i \cdot \hat{\bm s}_{i\alpha }
, \label{eqn_2ch_KLM}
\end{align}
where $c_{\bm{k}\alpha \sigma}$ is annihilation operator of the conduction electron with momentum $\bm k$, channel $\alpha =1,2$ and pseudo-spin $\sigma = \uparrow, \downarrow$.
The chemical potential has been set to zero in $\varepsilon_{\bm k}$.
The localized pseudo-spin at site $i$ is described by the spin 1/2 operator $\hat{\bm S}_i$, while 
$\hat{\bm s}_{{\rm c} i \alpha}$
represents the conduction pseudo-spin of channel $\alpha$ at site $i$.

In the single-impurity system, the ground state of the model has a residual entropy $\ln\sqrt 2$.   This non-Fermi liquid state is very fragile, and becomes unstable against certain perturbations.   The most relevant ones are
the followings \cite{affleck92}:
\begin{align}
V_1^z = h_1\hat{S}^z, \quad
V_2^z = h_2\hat{\bm S}\cdot (\hat{\bm s}_1-\hat{\bm s}_2),
\end{align}
where $V_1^z$ is the pseudo-Zeeman term with pseudo magnetic field $h_1$ that breaks the non-Kramers degeneracy, and $V_2^z$ breaks the equivalence of two channels.
Because of the SU(2) symmetry against the channel rotation, the operator in $V_2^z$ is regarded as the $z$-component of a vector operator $\bm\Psi$.  
In general the $\mu$-component $\Psi^\mu$ with $\mu=x,y,z$ is given by
\begin{align}
\Psi^\mu= \frac 12
\hat{\bm{S}} \cdot 
{\bm\sigma}_{\sigma\sigma'}
c_{\alpha\sigma}^\dagger 
\sigma^\mu_{\alpha\beta}
c_{\beta\sigma'}
\end{align}
Here and in the rest of the present section, 
we take the Einstein convention to omit summation symbols over repeated spin variables.
The $x$ and $y$ components of $\bm\Psi$ are equivalent to $\Psi^z$ with SU(2) symmetry, which we shall exploit later to study the superconducting order.

In the lattice system, the residual entropy is removed even without external perturbations such as $V_1^z$ and $V_2^z$.  Namely, spontaneous ordering corresponding to these operators takes place.  In contrast with the ordinary Kondo lattice that can stay without any order, the 2CKL must have some kind of order because of the remaining entropy otherwise.

\begin{figure}
\begin{center}
\includegraphics[width= 0.4\textwidth]{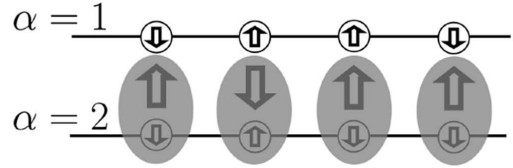}
\caption{
Illustration of the homogeneous composite order.
The arrows show the pseudospins that describe orbital
degrees of freedom, while $\alpha =1,2$ describe the real up- and down-spins, respectively.
}
\label{composite}
\end{center}
\end{figure}

The simplest order from non-Fermi liquid has the mean-field described by $V_1^z$. 
Since this kind of order is connected continuously with a simple multipole order of localized electrons, 
characteristics peculiar to two channels do not appear.
On the other hand, another order described by $V_2^z$ is more interesting since it
breaks the time-reversal symmetry 
even with zero moment: $\langle
\hat{\bm s}_1-\hat{\bm s}_2\rangle=0.$
Note that the pseudo-spin describes the non-magnetic degrees of freedom, while
the channel 1 and 2 correspond to the up- and down-spins, respectively.
Figure \ref{composite} illustrates the pattern obtained in Ref.\citen{hoshino2010} with the homogeneous order parameter $\Psi^z$.
Physically the order describes the breakdown of the channel symmetry;
the orbital Kondo insulator appears involving down-spin of conduction bands, while the up-spin conduction electrons make a Fermi liquid.  The different spatial distributions of magnetic moments 
means the broken time-reversal, although there is not net magnetization.
In this sense, the order parameter $\Psi^z$ represents a state with itinerant octupoles.
In another case where the non-Kramers degrees of freedom represents hexadecapoles, $\Psi^z$ produces an itinerant triakontadipole order.

We can also consider a situation where $\Psi^z$ forms a staggered order instead of the homogeneous one.
Such state is a candidate of the hidden order in 
URu$_2$Si$_2$, which is a subject of long-standing debate \cite{mydosh11}.

%%%%%%%%%%%%%
%%%%%%%%%%%%%

We show now that the composite operator $\Psi^z$ can be regarded as a part of the odd-frequency order of channel density, which is defined by
\begin{equation}
{\cal O}^z \equiv  \frac 1 N \sum_{i}
c_{i\alpha\sigma}^\dagger \sigma^z_{\alpha\beta} \dot{c}_{i\beta\sigma}
\end{equation}
where $\dot{c}_{i\alpha\sigma} = [{\cal H}, c_{i\alpha\sigma}]$ is a imaginary-time derivative of the annihilation operator for channel $\alpha=1,2$.
By calculating the commutator with the Hamiltonian, we obtain
\begin{align}
{\cal O}^z   & \equiv
{\cal O}^z_1+{\cal O}^z_2 =   
\frac{1}{N} \sum _{\bm{k}}
\varepsilon _{\bm{k}} 
c_{\bm{k}\alpha \sigma}^\dagger \sigma^z_{\alpha\beta}
c_{\bm{k}\beta \sigma} 
\nonumber \\&
+ \frac{J}{2N} \sum_{i} 
\bm{S}_i \cdot 
{\bm\sigma}_{\sigma\sigma'}
c_{i \alpha\sigma}^\dagger 
\sigma^z_{\alpha\beta}
c_{i \beta\sigma'},
\label{dot-c}
\end{align}
where the second term ${\cal O}^z_2$ defines $\Psi^z$ for the lattice as
${\cal O}^z_2 = J\Psi^z$.
The first term ${\cal O}^z_1$ represents the difference of kinetic energy between the channels.  The origin of difference is easily recognized from Fig.\ref{composite}.

\subsection{Virtual $f$ Electron Operators}
Let us consider the composite order in more intuitive terms.  For this purpose we  introduce  creation and annihilation operators of virtual $f$ electrons, 
although there is no charge degrees of freedom in the Kondo lattice
for real $f$ electrons.
Related ideas have been discussed with Majorana particles\cite{coleman94},
and for the Hubbard model \cite{mancini2004}.
The imaginary-time derivative $\dot{c}_{i\alpha\sigma}$ includes the contribution 
from the Kondo interaction ${\cal H}_{\rm ex}$, and is given by
\begin{align}
[{\cal H}_{\rm ex}, c_{i\alpha\sigma} ]
=
-\frac J 2 \hat{\bm S}_i \cdot \bm 
\sigma_{\sigma\sigma'} c_{i\alpha\sigma'} \equiv 
-\frac J 2 \tilde{f}_{i\alpha\sigma},
\end{align}
where $\tilde{f}_{i\alpha\sigma}$ has been introduced to represent annihilation of the virtual $f$ electron.
The merit of using 
the virtual $f$ electron is most simply seen in the ordinary Kondo lattice;  the heavy band can be understood as a result of hybridization between $c$ and virtual $f$ electrons.
Namely we rewrite the exchange term in a hybridization form:
\begin{align}
{\cal H}_{\rm ex} = \frac J 4 \sum_i (\tilde f^\dagger_{i\alpha\sigma} c_{i\alpha\sigma} + c^\dagger_{i\alpha\sigma} \tilde f_{i\alpha\sigma}).
\end{align}
Note that 
$\tilde{f}_{i\alpha\sigma}$ and
$\tilde{f}_{i\alpha\sigma}^\dagger$ do not satisfy  
the anti-commutation rule of ordinary fermions.
As a result, the width of the heavy band is not given by $J$, but by the 
Kondo temperature.

Using the virtual $f$ electrons, we can rewrite 
${\cal O}_2^z$ in eq.(\ref{dot-c}) as
\begin{align}
{\cal O}_2^z = 
\frac{J}{2N} \sum_{i}
c_{i \alpha\sigma}^\dagger 
\sigma^z_{\alpha\beta}
\tilde{f}_{i\beta\sigma}
= \frac{J}{2N} \sum_{i}
\tilde{f}_{i \alpha\sigma}^\dagger 
\sigma^z_{\alpha\beta}
c_{i\beta\sigma}.
\end{align}
Namely, the composite order describes 
a channel symmetry breaking involving virtual hybridization.

\subsection{Off-Diagonal Order of Itinerant-Localized Composites}

Let us proceed to off-diagonal orders relevant to superconductivity.
We start from a generalized pairing function
\begin{align}
\Phi_{\alpha\beta} (\tau)\equiv \langle c_\alpha c_\beta (\tau) \rangle
= \psi_0 +\tau\psi_{\rm odd}+ \tau^2\psi_2+\cdots,
\end{align}
where $\psi_0$ gives the ordinary pairing, while we obtain
$\psi_{\rm odd} = \langle c_\alpha \dot{c}_\beta  \rangle
$
by definition.
If we have $\psi_0=0$  and $\psi_{\rm odd} \neq 0$, there is no ordinary pairing amplitude, 
but the gauge symmetry is broken.
Such state is called the odd-frequency (OF) pairing.
Possible relevance of the OF pairing 
to real materials was first 
pointed out by Berezinskii for $^3$He \cite{berezinskii74}.
After the discovery of high-temperature superconductivity in cuprates,
the OF pairing has aroused broad interest 
\cite{emery92, balatsky92, abrahams93, emery93, coleman94,
jarrell97, anders02-2} 
as one of candidate mechanisms for high temperature superconductivity.

We now fully exploit the symmetry of the 2chKL, and show that the energy of the diagonal composite order is the same as the odd-frequency superconducting order, provided the two conduction bands are both half-filled.
Let us consider a bipartite lattice, and 
introduce a particle-hole (PH) transformation ${\cal P}_{\rm ph}$ that depends on sublattice indices A and B:
\begin{align*}
c_{{\rm A}\uparrow} & \rightarrow  c_{{\rm A}\downarrow}^\dagger, \quad
c_{{\rm A}\downarrow} \rightarrow -c_{{\rm A}\uparrow}^\dagger, \\
c_{{\rm B}\uparrow} & \rightarrow -c_{{\rm B}\downarrow}^\dagger, \quad
c_{{\rm B}\downarrow} \rightarrow  c_{{\rm B}\uparrow}^\dagger. 
\end{align*}
The transformation is compactly written as
\begin{align}
{\cal P}_{\rm ph}c_{i\sigma}{\cal P}_{\rm ph}^{-1} = 
\epsilon_{\sigma\sigma'}c_{i\sigma'}^\dagger \exp({\rm i}\bm Q\cdot\bm R_i)
\end{align}
with $\epsilon_{\sigma\sigma'} \equiv {\rm i}\sigma^y_{\sigma\sigma'}$, and 
$\bm Q$ being the vector corresponding to the staggered order.
The kinetic energy associated with the hopping between the nearest-neighbor sites remain the same after the PH transformation.  Moreover, the spin operators also remain the same.  The invariance can be seen from
\begin{align}
t c_{{\rm A}\downarrow}^\dagger c_{{\rm B}\downarrow} 
& \rightarrow 
 -t c_{{\rm A}\uparrow} c_{{\rm B}\uparrow}^\dagger
= t c_{{\rm B}\uparrow}^\dagger c_{{\rm A}\uparrow} , 
\label{kinetic}\\
c_{i\uparrow}^\dagger c_{i\downarrow} 
& \rightarrow
-c_{i\downarrow} c_{i\uparrow}^\dagger =  
c_{i\uparrow}^\dagger c_{i\downarrow}.   
\label{spin}
\end{align}

Let us now consider the component
\begin{align*}
\Psi^-  \equiv \Psi^x -i\Psi^y =
\frac 1{2N} \sum_i
\bm{S}_i \cdot {\bm \sigma}_{\sigma\sigma'}
c^\dagger_{i2\sigma}c_{i1\sigma'}
\end{align*}
which is energetically degenerate with $\Psi^z$ because of the SU(2) channel symmetry.
The crucial step is to make the PH transformation to $
\Psi^-$ only for the channel 2.  The result is given by
\begin{align*}
\Psi^-
&\rightarrow \frac 1{2N}\sum_i \exp({\rm i}\bm Q\cdot \bm R_i)
{\bm{S}}_i \cdot \left( 
\epsilon{\bm \sigma} \right)_{\sigma\sigma'}
c_{i2\sigma}
c_{i1\sigma'} \\
&\equiv \Phi(\bm Q), 
\end{align*}
where $\Phi(\bm Q) $ represents a
 composite superconducting order.
In analogy to eq.(\ref{dot-c}), 
$\Phi(\bm Q)\equiv {\cal O}_2(\bm Q)/J$ is a part corresponding to the odd-frequency order as given by
\begin{align}
{\cal O}(\bm Q) &=
\frac 1N\sum_i \exp({\rm i}\bm Q\cdot \bm R_i)
\epsilon_{\sigma\sigma'}
c_{i2\sigma}
\dot{c}_{i1\sigma'} \nonumber\\
&\equiv {\cal O}_1(\bm Q) +{\cal O}_2(\bm Q),
\label{Phi}
\end{align}
where we have decomposed ${\cal O}(\bm Q)$ as in the diagonal order ${\cal O}^z$.
The first part ${\cal O}_1(\bm Q)$ is given by
\begin{align}
{\cal O}_1(\bm Q) 
&= 
\frac {\epsilon_{\sigma\sigma'}} N
\sum_{\bm k} \varepsilon_{\bm k}
c_{\bm k 2\sigma}  c_{-\bm k-\bm Q, 1\sigma'}.
 \label{eqn_second_pair}
\end{align}
Note that $\langle{\cal O}_1(\bm Q)\rangle $ is analogous to the $\eta$-pairing \cite{yang89} in the sense of an even-frequency and staggered pairing.  The difference however is the presence of the factor $\varepsilon_{\bm k}$ with the $\bm k$-component.

Using the relation
$\langle c_{i2\sigma}\dot{c}_{i1\sigma'} \rangle = 
- \langle \dot{c}_{i2\sigma} c_{i1\sigma'} \rangle$
we can rewrite
\begin{align}
\langle \Phi (\bm Q) \rangle
&=
\frac 1{2N}\sum_i \exp({\rm i}\bm Q\cdot \bm R_i)
\epsilon_{\sigma\sigma'}
\epsilon_{\alpha\beta}
\langle
\tilde{f}_{i\alpha\sigma}
c_{i\beta\sigma'} \rangle ,
\label{Phi2}
\end{align}
where the virtual $f$-electron operator has been used.
Thus the composite superconducting order can be interpreted as the even-frequency pairing between $c$ and virtual $f$ electrons.  

Since the Hamiltonian is invariant against the PH transformation as shown in eqs.(\ref{kinetic}) and (\ref{spin}), the odd-frequency order characterized by $\langle\Phi\rangle$ has the same energy as the diagonal orders $\langle \Psi^\mu\rangle $ with $\mu=x,y,z$.
However the number of particles and holes are the same only for half-filling, and the degeneracy no longer remains off half-filling.  It has been shown \cite{hoshino2013} that the composite superconductor is the most stable for a wide density range between quarter- and half-fillings.
More details of the composite order are discussed elsewhere\cite{hoshino11,hoshino13,hoshino2013}.

\subsection{SO(5) Symmetry}
We have thus demonstrated by explicit calculation the five-fold degeneracy between the real diagonal order parameters from $\Psi^x,\Psi^y,\Psi^z$ and the off-diagonal ones from $\Phi+\Phi^\dagger, {\rm i}(\Phi-\Phi^\dagger)$.
The degeneracy is interpreted in terms of the SO(5) symmetry\cite{affleck92,demler-zhang2004}.  Namely, the order parameters constitute a basis of the vector representation of the SO(5) group. 
For the two-channel Kondo impurity, presence of an SO(5) symmetry, or equivalently the Sp(4) symmetry,  has long been pointed out \cite{affleck92}.  
In the context of high-$T_{\rm c}$ cuprates, extensive study of the approximate SO(5) symmetry
has been performed \cite{demler-zhang2004}.  

Instead of the AF state in cuprates,  we have in our case the homogeneous composite operator $\Psi^z$.  Correspondingly, the $d$-wave superconducting state in cuprates is replaced by the composite pairing operator $\Phi(\bm Q)$ with the pair momentum $\bm Q$.  
With this difference considered, it is straightforward to construct the Ginzburg-Landau phenomenology containing SO(5) composite order parameters.

\section{Antiferromagnetism and Lifshitz Transition}
\subsection{Localized vs Itinerant Antiferromagnetism}

Let us turn our attention to the Kramers system,
and assume that the occupation numbers $n_f$ of $f$ electrons (f hole in case of Yb) and 
$n_c$ of conduction electrons at each site satisfy the condition $n_c+n_f=2$. 
The following two extreme cases are assumed. In the first case, the energy band is composed only of $c$ electrons since $f$ electrons are strictly localized with $n_f=1$. Then the energy band is half-filled, and a metallic state is realized. 
On the contrary, if $f$ electrons also participate in the band, the electron number per unit cell becomes an even number $( =2)$.
In this second case, the valence band that mixes $f$ and $c$ electrons is completely filled, and the system is an insulator. 

In the intermediate case, the behavior of the system depends on temperature $T$; a metallic state is realized at high $T$ where $f$ electrons behave as localized, while insulating state is formed at low enough $T$.  Such insulating state at low $T$ is called the Kondo insulator.  The crucial difference from the ordinary insulator is that the magnitude of the energy gap depends strongly on $T$.
Description of the $T$-dependent energy gap is much beyond the scope of the ordinary band theory.

\begin{figure}
\begin{center}
\includegraphics[width= 0.4\textwidth]{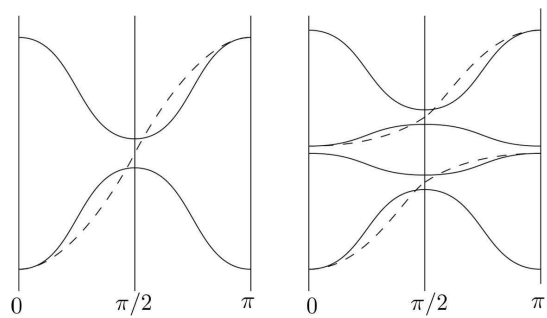}
\caption{
Illustration of one-dimensional energy bands of localized AFM (left), and itinerant AFM with Kondo effect (right).  The dashed lines show the energy bands in the paramagnetic state with $\pi$ being the boundary of the BZ.
}
\label{AF-kondo}\end{center}
\end{figure}

Let us consider the AFM state of the system.  Without Kondo effect, the single conduction band is split into two by the new periodic potential with doubled periodicity.  The new Brillouin zone (BZ) becomes half of the original one. Figure \ref{AF-kondo} (left) illustrates the situation.  Because the Fermi level 
sits inside the newly formed gap, the system becomes an insulator.
We refer to the state as the localized antiferromagnet (LAF), which corresponds to Mott insulator of $f$ electrons. 

Depending on combination of the parameters, the Kondo insulator undergoes an AFM ordering, which is called the Kondo antiferromagnet (KAF).  The resultant band structure is illustrated in Fig.\ref{AF-kondo} (right). 
We may ask how these two kinds of insulators,
LAF and KAF, are connected to each other.
To study this problem it is convenient to consider the specific model called the Kondo-Heisenberg lattice, which is given by
\begin{align}
{\cal H} = \sum_{\bm{k} \sigma} \varepsilon_{\bm k} c_{\bm k \sigma}^\dagger c_{\bm k \sigma}
\hspace{-0.5mm}+\hspace{-0.5mm}
 J  \sum_i \bm{S}_i \cdot \bm{s}_i
\hspace{-0.5mm}+\hspace{-0.5mm}
 \frac{J_{\rm H}}{z} \sum_{(ij)} \bm{S}_i \cdot \bm{S}_j,
\end{align}
where $J_{\rm H}$ corresponds to $\Delta$ in the two-site system shown in Fig.\ref{4spins}, but involves the number $z$ of neighboring sites for normalizing the N\'{e}el temperature $T_{\rm N}$.

Let us start with 
the insulating ground state 
for vanishing value of $J_{\rm H}/J$ and a small bandwidth $2D$ of the conduction band as compared with $J$.
Then each site forms a Kondo singlet with negligible intersite interaction.  
In the opposite limit of $J \ll J_{\rm H},D$, 
the ground state becomes a localized AFM with slight interaction with conduction electrons.
As $J_{\rm H}$ increases from zero, and becomes comparable to $J$ (or $T_{\rm K}$ for large band-width), Kondo singlets begin to compete the stability with the AFM.
The AFM may be formed around the region with significant Kondo effect as well.  This is the picture of the KAF.
One may ask whether the change from KAF to LAF is continuous, or involves a phase transition.
According to the result for the two-site system shown in Fig.\ref{4spins}, the ground state changes smoothly unless the particular condition $\Delta=2J=2I$ is satisfied.

%%%%%%%%%%%%%
%%%%%%%%%%%%%

\subsection{Topology of Fermi Surface}

\begin{figure}
\includegraphics[width= 0.5\textwidth]{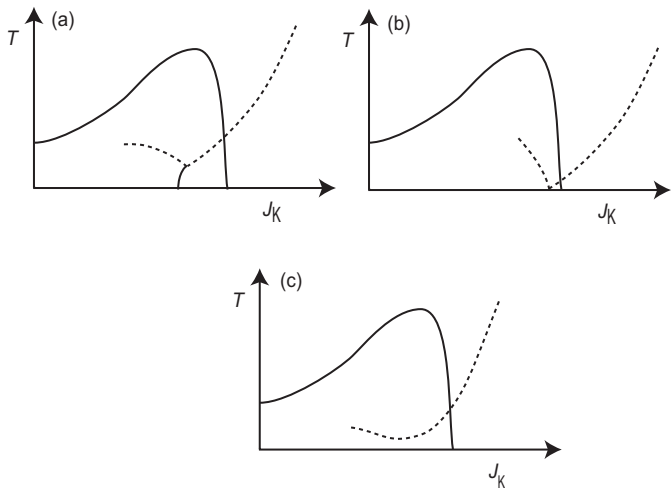}
\caption{
Schematic phase diagram of the Kondo-Heisenberg lattice. (a) Generic case with the first-order Lifshitz transition terminating at the critical point.  The dashed lines show the characteristic temperature as probed by the peak in the resistivity. (b) Special case where the critical temperature of the Lifshitz transition is almost zero, and occurs near the AFM quantum critical point.
(c) Case without the Lifshitz transition. The characteristic temperature changes continuously across the itinerant and localized AFM.
This occurs with insulating ground state.  Another possibility is discussed in the text.
}
\label{AF-phase}\end{figure}

Infinitesimal deviation from $n_c=1$ corresponds to carrier doping into the AFM insulating state of the Kondo-Heisenberg lattice.
The Fermi 
level
 is near the top of the valence band if $n_c\lesssim 1$.
The location of the Fermi surface is different whether the reference state is LAF or KAF.  This situation is clearly seen in Fig.\ref{AF-kondo}.
In the LAF, the top of the valence band is located in the boundary $(k=\pi/2)$ of the AF (super-lattice) BZ. 
On the other hand, in the KAF it is located in the center $(k=0,\pi)$ of the AF BZ. 
On the other hand, it may naturally be guessed that the energy of the system changes only infinitesimally for infinitesimal doping.
Since the location of the Fermi surface 
is different, the change is generally discontinuous.
Such change of the Fermi surface is called the Lifshitz transition.
Figure \ref{AF-phase} illustrates possible phase diagrams including the Lifshitz transition.

%%%%%%%%%%%%%%
%%%%%%%%%%%%%%

We have the following options to avoid the first-order transition:\\
(i) In the intermediate region between itinerant and localized limits, there occurs multiple maxima in the valence band or minima in the conduction band.\\
(ii) The entire valence band becomes flat at the transition so that the change of the Fermi surface can occur continuously.

The possibility (i) was emphasized in Ref.\citen{martin10}.  Then the carriers in one pocket move to the other continuously by changing the ratio of population in the two pockets. 
The possibility (ii) 
corresponds to the quantum critical Lifshitz transition, which
brings about various anomalies as will be discussed later.
It is shown numerically in ref.\citen{hoshino13prl} that the local susceptibility is greatly enhanced at the Lifshitz transition, which can be understood by partial flattening of the valence band.  

On the other hand, 
if the discontinuity of the Lifshitz transition
is large, the AFM transition occurs simultaneously.
Then we observe the first-order transition from the AFM of localized $f$ electrons to paramagnetic heavy electrons with a large Fermi surface.

In actual systems with finite doping, such transition should accompany more or less the change of $n_f$ that is related to the valency.  We note, however, the transition is possible also in the Kondo-Heisenberg lattice with $n_f$ fixed to unity.  This result implies
 that the valence change is a consequence of the transition rather than its cause.

Persistence of Kondo effect into the AFM phase in CeIn$_3$ has been
probed by optical conductivity \cite{kimura}.
Namely in the pressure range below the collapse of the AFM, 
new absorption emerges that seems to originate from Kondo effect.

The AFM in the presence of Kondo effect seems relevant also 
to Kondo insulators 
CeT$_2$Al$_{10} $ (T=Ru, Os). \cite{sera-order,taka-order} 
For instance, the magnetic susceptibility $\chi (T)$ below $T_{\rm N}$,
which is $27$ K in CeRu$_2$Al$_{10}$,  
decreases along all direction of the magnetic field \cite{sera}.
This behavior cannot be understood from the molecular field theory of AFM, but shows that Kondo effect is predominant.
On the other hand, if Rh is doped slightly ($\sim 5 \%$), 
the behavior of $\chi (T)$
becomes almost the same as that of usual antiferromagnetism as in NdRu$_2$Al$_{10}$ etc \cite{kondo13}.
We interpret that
CeRu$_2$Al$_{10} $ is located near the limit from KAF to LAF, and 
the  Lifshitz transition is triggered by slight change of environment.
In neutron inelastic scattering \cite{mignot},
strong excitation intensity of about 8 meV is observed only in the ordered phase. 
Furthermore, in the optical conductivity, 
a new structure of about 20 meV emerges in the ordered phase \cite{kimuraCeRuAl}. 

Finally we mention a variational calculation in two-dimensional Kondo lattice \cite{watanabe-ogata2007}, 
which has derived a change of the Fermi surface in the AF phase. 
The LAF, however, cannot be derived by this approach since
the Gutzwiller wave function in the variational approach cannot describe the RKKY interaction.
Description of enhanced local susceptibility\cite{hoshino13prl} is also beyond the scope of the Gutzwiller wave function.

\subsection{AFM Quantum Critical Point in Low-Dimensional Systems}

Since the Kondo effect remains robust at the AFM QCP,  we may regard the AFM on the small $J$ side as itinerant, or SDW.
Our result becomes exact in the limit of infinite dimensions, and
agrees with the renormalization theory \cite{hertz,millis,moriya} of the QCP for three dimensions.
The itinerant feature remains in the limit of Kondo insulator, as seen from the locations of extrema in conduction and valence bands.

However, the DMFT is no longer reliable in two-dimensional systems. 
It is known that the coupling between electrons and spin fluctuations becomes stronger \cite{hertz,millis} according to the Ginzburg-Landau-Wilson theory. 
Elaborate numerical calculation \cite{martin10},
using a scheme called the DCA, 
has been made for the two-dimensional Kondo lattice.  
The result shows that the magnetism arising at the QCP is regarded as an itinerant AFM.

There is a possibility that two-dimensional character survives down 
to sufficiently low temperatures, although eventual crossover to 
three dimensions is unavoidable.
In such a case, substantial entropy may remain as in the non-Fermi liquid of 2chKL. 
The entropy is removed by some nonmagnetic orders involving superconductivity.
Systems such as CeRhIn$_5$ have a quasi-two-dimensional character in the electronic structure, and its superconductivity can be qualitatively understood along this line.
 
Experimentally, $T_N$ of YbRh$_2$Si$_2$ decreases with application of magnetic field, and becomes zero for $B = B_{\rm c}\sim 60$mT.  Around this magnetic field, the characteristic energy of the system becomes tiny, and the Hall coefficient shows a rapid change. 
It has been proposed \cite{gegenwart} that Kondo effect is effective only in the paramagnetic phase, and $f$ electrons for $B<B_{\rm c}$ are localized.
This scenario is difficult to reconcile with 
ARPES results that probe the $f$-electron band already at $T\sim 10$ K without magnetic field \cite{vyalikh10, mo12}.
The mysterious behavior of spin fluctuations in YbRh$_2$Si$_2$ needs more study \cite{lohneysen2007} before complete solution of the problem.

\section{Summary and Conclusion}

We have discussed electronic orders in heavy electrons, and 
the change of the Fermi surface which occurs inside the AF order. 
In non-Kramers systems described by a singlet-triplet Kondo lattice, we have demonstrated
the scalar order with
staggered sublattices of crystalline-electric-field (CEF) and Kondo singlets, which provides natural interpretation
of the observed order in PrFe$_4$P$_{12}$.
In the two-channel Kondo lattice, 
An odd-frequency order is equivalent to an even-frequency composite order.
Furthermore the composite order is more intuitively understood in terms of virtual $f$ electrons.
With half-filling of each conduction band, 
the odd-frequency pairing is degenerate with
the 
channel order involving virtual $f$ electrons.
The degeneracy is interpreted as a consequence of the hidden SO(5) symmetry as in the case of a two-channel Kondo impurity. 

For the ordinary Kondo lattice 
we have taken insulating ground states as reference, and characterized the itinerant vs localized magnetism in terms of location of the Fermi surface.
By adding intersite exchange and with finite doping,
we find that the Lifshitz transition occurs in general separately from the quantum critical point of the antiferromagnetic order.

%acknowledgement
The authors thank J. Otsuki for useful conversations.
This work was supported partly by 
MEXT KAKENHI ``Heavy Electrons" Grant Number 20102008,  
and by JSPS KAKENHI Grant Number 24340072.


\begin{thebibliography}{99}

\bibitem{kuramoto85} Y. Kuramoto: {\it Theory of Heavy Fermions and Valence Fluctuations}, Eds. T. Kasuya and T. Saso (Springer, 1985) p.152.

\bibitem{georges96} For a review, see A. Georges, G. Kotliar, W. Krauth and M. J. Rozenberg: Rev. Mod. Phys. {\bf 68} (1996) 13.

\bibitem{rubtsov04} A. N. Rubtsov and A. I. Lichtenstein: JETP Lett. {\bf 80} (2004) 67.

%\bibitem{werner06} P. Werner, A. Comanac, L. Medici, M. Troyer, and A. J. Millis: Phys. Rev. Lett. {\bf 97} (2006) 076405.

\bibitem{otsuki07} J. Otsuki, H. Kusunose, P. Werner. and Y. Kuramoto: J. Phys. Soc. Jpn. {\bf 76} (2007) 1147076.

\bibitem{gull11}  For a review, see E. Gull, A. J. Millis, A. I. Lichtenstein, A. N. Rubtsov, M. Troyer and P. Werner: Rev. Mod. Phys. {\bf 83} (2011) 349.

\bibitem{wilson75} K. G. Wilson:  Rev. Mod. Phys. {\bf 47} (1975) 773. 
\bibitem{bulla} R. Bulla, T. A. Costi, and T. Pruschke:  Rev. Mod. Phys. {\bf 80} (2008) 395. 


\bibitem{gegenwart}
P. Gegenwart, Q. Si, and F. Steglich: Nat. Phys. {\bf 4} (2008) 186.

\bibitem{lohneysen2007}
H.v. L\"{o}hneysen, A. Rosch, M. Vojta, and P. W\"{o}lfle:
 Rev. Mod. Phys. {\bf 79} (2007) 1015. 

\bibitem{otsuki09} J. Otsuki, H. Kusunose and Y. Kuramoto: Phys. Rev. Lett. {\bf 102} (2009) 017202.


\bibitem{hoshino2010}
S. Hoshino, J. Otsuki, and Y. Kuramoto:
J. Phys. Soc. Jpn. {\bf 79} (2010) 074720.

\bibitem{kuramoto2011}
Y. Kuramoto, S. Hoshino and J. Otsuki: 
J. Phys. Soc. Jpn. {\bf 80} Supplement A (2011) SA018.

\bibitem{cox87} D. L. Cox: Phys. Rev. Lett. {\bf 59} (1987) 1240.
\bibitem{cox98} For a review, see D. L. Cox and A. Zawadowski: Adv. Phys. {\bf 47} (1998) 599.


\bibitem{hoshino11} S. Hoshino, J. Otsuki and Y. Kuramoto: Phys. Rev. Lett. {\bf 107} (2011) 247202.

\bibitem{hoshino13} S. Hoshino, J. Otsuki and Y. Kuramoto: J. Phys. Soc. Jpn. {\bf 82} (2013) 044707.
\bibitem{hoshino2013} S. Hoshino and Y. Kuramoto: arXiv:1309.5719 (2013).


\bibitem{rasul89}
J.W. Rasul and P. Schlottmann: Phys. Rev. Lett. {\bf 62} (1989) 1701.

\bibitem{affleck92-2}
I. Affleck and W. W. Ludwig: Phys. Rev. Lett. {\bf 68} (1992) 1046.

\bibitem{tanida06}
H. Tanida, H. Suzuki, S. Takagi, H. Onodera, and K. Tanigaki: J. Phys. Soc. Jpn. {\bf 75} (2006) 073705.
\bibitem{yatskar96} A. Yatskar, W. P. Beyermann, R. Movshovich, and P. C. Canfield: Phys. Rev. Lett: {\bf 77} (1996) 3637.
\bibitem{onimaru}
T. Onimaru, T. Sakakibara, A. Harita, T. Tayama, D. Aoki, and Y. Onuki: 
J. Phys. Soc. Jpn. {\bf 73} (2004) 2377.
\bibitem{sakai}
A. Sakai and S. Nakatsuji, J. Phys. Soc. Jpn. {\bf 80} (2011) 063701.


\bibitem{nozieres80} Ph. Nozi\`{e}res and A. Blandin: J. Physique {\bf 41} (1980) 193.


\bibitem{ott83} H. R. Ott, H. Rudigier, Z. Fisk and J. I. Smith: Phys. Rev. Lett. {\bf 50} (1983) 1595.

\bibitem{torikachvili87} M. S. Torikachvili, J. W. Chen, Y. Dalichaouch, R. P. Guertin, M. W. McElfresh, C. Rossel, and M. B. Maple: Phys. Rev. B {\bf 87} (1987) 8660.

\bibitem{hao03} L. Hao, K. Iwasa, M. Nakajima, D. Kawana, K. Kuwahara, M. Kohgi, H. Sugawara, T. D. Matsuda, Y. Aoki, and H. Sato: Acta Phys. Pol. B {\bf 34} (2003) 1113.

\bibitem{matsuda02} 
T. D. Matsuda, S. R. Saha, T. Namiki, H. Sugawara, Y. Aoki and H. Sato: 
J. Phys. Soc. Jpn. {\bf 71} Supplement (2002)  246.

\bibitem{tayama07} T. Tayama, Y. Isobe, T. Sakakibara, H. Sugawara, Y. Aoki, and H. Sato: J. Phys. Soc. Jpn. {\bf 76} (2007) 083702.
\bibitem{hidaka05} H. Hidaka, I. Ando, H. Kotegawa, T. Kobayashi, H. Harima, M. Kobayashi, H. Sugawara, and H. Sato: J. Phys. Soc. Jpn. {\bf 71} (2005) 073102.


\bibitem{jarrell96} M. Jarrell, H. Pang, D. L. Cox, and K. H. Luk: Phys. Rev. Lett. {\bf 77} (1996) 1612.

\bibitem{affleck92} I. Affleck, A. W. W. Ludwig, H.-B. Pang and D. L. Cox: Phys. Rev. B {\bf 45} (1992) 7918.

\bibitem{mydosh11} J. A. Mydosh and P. M. Oppeneer: Rev. Mod. Phys. {\bf 83} (2011) 1301.

\bibitem{coleman94} P. Coleman, E. Miranda and A. Tsvelik: Phys. Rev. B {\bf 49} (1994) 8955.

\bibitem{mancini2004} 
F. Mancini and A. Avella: Adv. in Phys. {\bf 53} (2004) 537.


\bibitem{berezinskii74} V. L. Berezinskii: Pis'ma Zh. Eksp. Teor. Fiz. {\bf 20} (1974) 628 [JETP Lett. {\bf 20} (1974) 287].

\bibitem{emery92} V. J. Emergy and S. Kivelson: Phys. Rev. B {\bf 46} (1992) 10812.
\bibitem{balatsky92} A. Balatsky and E. Abrahams: Phys. Rev. B {\bf 45} (1992) 13125.
\bibitem{emery93} V. J. Emery and S. A. Kivelson: Phys. Rev. Lett. {\bf 71} (1993) 3701.
\bibitem{abrahams93} E. Abrahams, A. Balatsky, J. R. Schrieffer and P. B. Allen: Phys. Rev. B {\bf 47} (1993) 513.
\bibitem{jarrell97} M. Jarrell, H. Pang and D. L. Cox: Phys. Rev. Lett. {\bf 78} (1997) 1996.
\bibitem{anders02-2} F. B. Anders: Eur. Phys. J. B {\bf 28} (2002) 9.


\bibitem{yang89} C. N. Yang: Phys. Rev. Lett. {\bf 63} (1989) 2144.


\bibitem{demler-zhang2004} E. Demler, W. Hanke and S.-C. Zhang: Rev. Mod. Phys. {\bf 76} 909 (2004). 


\bibitem{martin10} L. C. Martin, M. Bercx, and F. F. Assaad: {\bf 82} (2010) 245105.
\bibitem{hoshino13prl} 
S. Hoshino and Y. Kuramoto: Phys. Rev. Lett. {\bf 111} (2013) 026401.

\bibitem{kimura}
T. Iizuka, T. Mizuno, B. Min, Y. Kwon, and S. Kimura: J. Phys. Soc. Jpn. {\bf 81} (2012) 043703.

\bibitem{sera-order}
J. Robert, J-M. Mignot, G. Andre, T. Nishioka, R. Kobayashi, M. M., H. Tanida, D. Tanaka, and M. Sera
Phys. Rev. B {\bf 82}  (2010) 100404(R) 

\bibitem{taka-order}
D. D. Khalyavin, A. D. Hillier, D. T. Adroja, A. M. Strydom, P. Manuel, L. C. Chapon, P. Peratheepan, K. Knight, P. Deen, C. Ritter, Y. Muro, and T. Takabatake: 
Phys. Rev. B {\bf 82}  (2010) 100405(R) 

%\bibitem{takabatake}
%Y. Muro, K. Motoya, Y. Saiga and T. Takabatake: 
% J.Phys.Soc.Jpn. {\bf 78}  (2009) 083707.

\bibitem{sera}
H. Tanida, Y. Nonaka, D. Tanaka, M. Sera, Y. Kawamura, Y. Uwatoko, T. Nishioka, and M. Matsumura:Phys. Rev. B {\bf 85} (2012) 205208.

\bibitem{kondo13} A. Kondo, K. Kindo, K. Kunimori, H. Nohara, H. Tanida, M. Sera, R. Kobayashi, T. Nishioka, and M. Matsumura: J. Phys. Soc. Jpn. {\bf 82} (2013) 054709.

\bibitem{kobayashi13} R. Kobayashi, Y. Ogane, D. Hirai, T. Nishioka, M. Matsumura, Y. Kawamura, K. Matsubayashi, Y. Uwatoko, H. Tanida, and M. Sera: J. Phys. Soc. Jpn. {\bf 82} (2013) 093702.

\bibitem{mignot} J. Robert, J. Mignot, S. Petit, P. Steffens, T. Nishioka, R. Kobayashi, M. Matsumura, H. Tanida, D. Tanaka, and M. Sera: Phys. Rev. Lett. {\bf 109} (2012) 267208.


\bibitem{kimuraCeRuAl}
S. Kimura, T. Iizuka, H. Miyazaki, T. Hajiri, M. Matsunami, T. Mori, A. Irizawa, Y. Muro, J. Kajino, and T. Takabatakel: Phys. Rev. B {\bf 84} (2011) 165125.


\bibitem{watanabe-ogata2007}
H. Watanabe and M. Ogata: Phys. Rev. Lett. {\bf 99} (2007) 136401.

\bibitem{hertz}
J. A. Hertz: Phys. Rev. B {\bf 14} (1976) 1165.
\bibitem{millis}
A. J. Millis: Phys. Rev. B {\bf 48} (1993) 7183.
\bibitem{moriya}
T. Moriya and T. Takimoto: J. Phys. Soc. Jpn. {\bf 64} (1995) 960.

\bibitem{vyalikh10} D. V. Vyalikh, S. Danzenb\"{a}cher, Yu. Kucherenko, K. Kummer, C. Krellner, C. Geibel, M. G. Holder, T. K. Kim, C. Laubschat, M. Shi, L. Patthey, R. Follath, and S. L. Molodtsov: Phys. Rev. Lett. {\bf 105} (2010) 237601.
\bibitem{mo12} S.-K. Mo, W. S. Lee, F. Schmitt, Y. L. Chen, D. H. Lu, C. Capan, D. J. Kim, Z. Fisk, C.-Q. Zhang, Z. Hussain, and Z.-X. Shen: Phys. Rev. B {\bf 85} (2012) 241103(R).


\end{thebibliography}
\end{document}